\documentclass{article}

\PassOptionsToPackage{numbers, compress}{natbib}

\usepackage[preprint]{neurips_2020_ml4ps}

\usepackage[utf8]{inputenc} 
\usepackage[T1]{fontenc}    
\usepackage{hyperref}       
\usepackage{url}            
\usepackage{amsmath}
\usepackage{amssymb}
\usepackage{booktabs}       
\usepackage{amsfonts}       
\usepackage{nicefrac}       
\usepackage{microtype}      
\usepackage[pdftex]{graphicx}
\usepackage{xcolor}
\usepackage{mathtools}

\usepackage{wrapfig}
\usepackage[ruled,vlined]{algorithm2e}

\usepackage[capbesideposition=outside,capbesidesep=quad]{floatrow}

\newcommand{\bx}{\mathbf{x}}
\newcommand{\bz}{\mathbf{z}}

\newcommand{\btheta}{\boldsymbol{\theta}}
\newcommand{\bfeta}{\boldsymbol{\eta}}
\newcommand{\bvartheta}{\boldsymbol{\vartheta}}
\newcommand{\bg}{\mathbf{g}}

\newcommand{\tr}{\tilde{r}}
\newcommand{\tp}{\tilde{p}}

\newcommand{\pois}{\text{Pois}}

\title{Simulation-efficient marginal posterior estimation with \textit{swyft}: stop wasting your precious time}

\author{%
  Benjamin Kurt Miller \\
  AMLab \\
  Informatics Institute \\
  Gravitation Astroparticle Physics Amsterdam (GRAPPA) \\
  Institute for Theoretical Physics Amsterdam\\
  University of Amsterdam, the Netherlands \\
  \texttt{b.k.miller@uva.nl}
  \And
  Alex Cole \\
  Gravitation Astroparticle Physics Amsterdam (GRAPPA) \\
  Institute for Theoretical Physics Amsterdam \\
  University of Amsterdam, the Netherlands \\
  \texttt{a.e.cole@uva.nl}
  \And
  Gilles Louppe \\
  Montefiore Institute \\
  University of Liège, Belgium \\
  \texttt{g.louppe@uliege.be} \\
  \And
  Christoph Weniger \\
  Gravitation Astroparticle Physics Amsterdam (GRAPPA) \\
  Institute for Theoretical Physics Amsterdam\\
  University of Amsterdam, The Netherlands \\
  \texttt{c.weniger@uva.nl}
}

\begin{document}

\maketitle

\begin{abstract}
  We present algorithms (a) for nested neural likelihood-to-evidence ratio estimation, and (b) for simulation reuse via an inhomogeneous Poisson point process cache of parameters and corresponding simulations.  Together, these algorithms enable automatic and extremely simulator efficient estimation of marginal and joint posteriors.  The algorithms are applicable to a wide range of physics and astronomy problems and typically offer an order of magnitude better simulator efficiency than traditional likelihood-based sampling methods.  Our approach is an example of likelihood-free inference, thus it is also applicable to simulators which do not offer a tractable likelihood function. Simulator runs are never rejected and can be automatically reused in future analysis. As functional prototype implementation we provide the open-source software package \emph{swyft}\footnote{\emph{swyft} is located at \href{https://github.com/undark-lab/swyft}{https://github.com/undark-lab/swyft}.}.
\end{abstract}

\section{Introduction}

Parametric stochastic simulators are ubiquitous in the physical sciences \cite{Banik_2018, Bartels_2016, Rodr_guez_Puebla_2016}.  However, performing parameter inference based on simulator runs using Markov chain Monte Carlo is inconvenient or even impossible if the model parameter space is large or the likelihood function is intractable.  This problem is addressed by so-called likelihood-free inference \cite{sisson2018handbook} or simulation-based inference \cite{papamakarios2019sequential,greenberg2019automatic,Hermans2019,Cranmer2020,Durkan2020} techniques. 
Among those, sequential neural ratio estimation based on amortized approximate likelihood-to-evidence ratios (SNRE-AALR)~\cite{Hermans2019} is closest to our method.

We propose \emph{Nested Ratio Estimation} (NRE), which approximates the likelihood-to-evidence ratio in a sequence of rounds. Loosely inspired by the contour sorting method of nested sampling \cite{Skilling2006, Feroz2008, Handley2015}, the scheme alternates between sampling from a constrained prior and estimating likelihood-to-evidence ratios. It allows for efficient estimation of any marginal posteriors of interest. Furthermore, we propose an algorithm that we call \emph{iP3 sample caching}, which facilitates simulator efficiency by automatizing the reuse of previous simulator runs through resampling of cached simulations.

The primary use case for these algorithms is the calculation of arbitrary, low-dimensional marginal posteriors, typically in one or two dimensions. In physics and astronomy, such marginals serve as the basis for scientific conclusions by constraining individual model parameters within uncertainty bounds. 
We implement a multi-target training regime where all marginal posteriors of interest can be learned simultaneously.  We find that learning is simplified when one calculates each marginal distribution directly rather than computing the full joint posterior and marginalizing numerically.  
Furthermore, the method facilitates effortless marginalization over arbitrary numbers of nuisance parameters, increasing its utility in high-dimensional parameter regimes--even to simulators with a tractable, yet high-dimensional, likelihood \cite{lensing}.

\section{Proposed methods}

\paragraph{Nested Ratio Estimation (NRE).} 
We operate in the context of simulation-based inference where our simulator $\bg$ is a nonlinear function mapping a vector of parameters $\btheta= (\theta_{1}, \dots, \theta_{d}) \in \mathbb{R}^{d}$ and a stochastic latent state $\bz$ to an observation $\bx = \bg(\btheta, \bz)$. The likelihood function is therefore $p(\bx \vert \btheta) = \int \delta(\bx - \bg(\btheta, \bz)) \, p(\bz \vert \btheta) \, d\bz$, with $\delta(\cdot)$ denoting the Dirac delta.  Consider a factored prior $p(\btheta) = p_1(\theta_{1}) \cdots p_d(\theta_{d})$ over the parameters, the joint posterior is given via Bayes' rule as $p(\btheta|\bx) = p(\bx|\btheta)p(\btheta)/p(\bx) $, where $p(\bx)$ is the evidence.

Our goal is to compute the marginal posterior, $p(\bvartheta \vert \bx)$, where $\bvartheta$ are the parameters of interest.  We denote all other parameters by $\bfeta$, such that $\btheta = (\bvartheta, \bfeta)$. 
The marginal posterior is obtained from the joint distribution $p(\bvartheta, \bfeta|\bx) \equiv p(\btheta|\bx)$ by integrating over all components of $\bfeta$,
\begin{equation}
\label{eqn:post}
p(\bvartheta \vert \bx)  \equiv \int p(\bvartheta, \bfeta | \bx) d\bfeta
= \int \frac{p(\bx | \bvartheta, \bfeta)}{p(\bx)}  
p(\btheta) 
d\bfeta
= \frac{p(\bx|\bvartheta)}{p(\bx)}p(\bvartheta)\;,
\end{equation}
where we used Bayes' rule and defined the marginal likelihood $p(\bx|\bvartheta)$ in the last step.

Just like in SNRE-AALR, we focus on a specific observation of interest, $\bx_0$.  
Only parameter values $\btheta$ 
that could have plausibly generated observation $\bx_0$
will significantly contribute to the integrals in Eq.~\eqref{eqn:post}. For implausible values the likelihood $p(\bx_0|\btheta)$ will be negligible.  We denote priors that are suitably constrained to plausible parameter values by $\tp(\theta_1, \dots, \theta_d)$. Similarly, $\tilde{\square}$ indicates quantities $\square$ that are calculated using the constrained prior. Therefore, a judiciously chosen constrained prior, accurately approximates the marginal posterior in place of our true prior beliefs,
\begin{equation}
p(\bvartheta | \bx_0) =  
\frac{p(\bx_0|\bvartheta)}{p(\bx_0)} p(\bvartheta) \simeq
\frac{\tp(\bx_0|\bvartheta)}{\tp(\bx_0)} \tp(\bvartheta)\;.
\end{equation}
The increased probability that constrained priors assign to the plausible parameter region cancels when dividing by the constrained evidence $\tilde p(\bx)$. We define the marginal likelihood-to-evidence ratio
\begin{equation}
	\label{eqn:likelihood_ratio}
	\tr(\bx, \bvartheta) 
	\equiv \frac{\tp(\bx \vert \bvartheta)}{\tp(\bx)} 
	= \frac{\tp(\bx, \bvartheta)}{\tp(\bx) \tp(\bvartheta)} 
	= \frac{\tp(\bvartheta \vert\bx )}{\tp(\bvartheta)}\;,
\end{equation}
which is sufficient to evaluate the marginal posterior in Eq.~\eqref{eqn:post}, and which we will now estimate.
Under the assumption of equal class population, it is known \cite{Hermans2019, Cranmer2015} that one can recover density ratios using binary classification to distinguish between samples from two distributions.  Our binary classification problem is to distinguish positive samples,
$(\bx, \bvartheta) \sim \tp(\bx, \bvartheta) = p(\bx \vert \bvartheta) \tp(\bvartheta)$, drawn jointly, and negative samples, $(\bx, \bvartheta) \sim \tp(\bx) \tp(\bvartheta)$, drawn marginally. 
The binary classifier $\sigma(f_{\phi}(\bx, \bvartheta))$ performs optimally when $f_{\phi}(\bx, \bvartheta) = \log \tr(\bx, \bvartheta)$, where $\sigma(\cdot)$ is the sigmoid function and $f_{\phi}$ is a neural network parameterized by $\phi$.  The associated binary cross-entropy loss function \cite{Hermans2019} used to train the ratio $\tr(\bvartheta, \bx_0)$ via stochastic gradient descent is given by 
\begin{equation}
    \ell = -\int \left[ \tp(\bx|\bvartheta)\tp(\bvartheta) \ln \sigma(f_\phi(\bx, \bvartheta)) + \tp(\bx)\tp(\bvartheta) \ln \sigma(-f_\phi(\bx,\bvartheta)) \right] d\bx\, d\bvartheta\;.
\end{equation}

\emph{We propose to iteratively improve marginal posterior estimates in $R$ rounds by employing posterior estimates from previous rounds to define constrained priors.}
In each round $r$, we estimate \emph{all} 1-dim marginal posteriors, using $d$ instances of the above marginal likelihood-to-evidence ratio estimation in parallel by setting $\bvartheta = (\theta_i)$ for $i=1, \dots, d$. 
To this end, we utilize the factorized constrained prior, $\tp^{(r)}(\boldsymbol\theta) = \tp^{(r)}_1(\theta_1)\cdots\tp^{(r)}_d(\theta_d)$, which is defined recursively by a cutoff criterion,
\begin{equation}
    \tp^{(r)}_i(\theta_{i}) 
    \propto 
    p_i(\theta_{i}) \Theta_{H} \left[ \frac{\tr^{(r-1)}_i(\bx, \theta_{i})}{\max_{\theta_{i}} \tr^{(r-1)}_i(\bx, \theta_{i})} - \epsilon \right],
    \label{eqn:it}
\end{equation}
where $\Theta_{H}$ denotes the Heaviside step function and $\epsilon$ denotes the minimum likelihood-ratio which passes through the threshold. We use $\tp^{(1)}(\btheta) = p(\btheta)$ as an initial prior in the iterative scheme.

In every round, each 1-dim posterior approximates a marginalization of the same underlying constrained posterior, allowing us to effectively reuse training data and train efficiently in a multi-target regime. The inference network is therefore divided into a featurizer $\mathbf{F}(\bx)$ with shared parameters and a set of $d$ independent Multi-layer Perceptons $\{\textrm{MLP}_i(\cdot, \cdot)\}_{i=1}^{d}$ which estimate individual 1-dim marginal posterior-to-prior ratios and do not share parameters, such that 
$f_{\phi, i}(\bx, \theta_i) = \textrm{MLP}_i(\mathbf{F}(\bx), \theta_i)$. We estimate every 1-dim marginal simultaneously by concatenating the output of the classifiers such that $\mathbf{f}_{\phi} = (f_{\phi, 1}, \dots, f_{\phi, d})$.

This technique is valid as long as the excluded prior regions do not significantly affect the integrals in Eq.~\eqref{eqn:post}.  For uncorrelated parameters, a sufficient criterion is that the impact on the marginal posteriors is small, which we guarantee through the iteration criterion Eq.~\eqref{eqn:it}.  In the case of a very large number of strongly correlated parameters the algorithm can inadvertently cut away tails of the marginal posteriors. Decreasing $\epsilon$ mitigates this effect. Discussion is left for future study. 

With this design, the posteriors from the final round can be used as an approximation of the true 1-dim marginal posteriors, $\tp^{(R)}(\theta_i \vert \bx_{0}) \approx p(\theta_i\vert \bx_{0})$, while previous rounds were used to iteratively focus on relevant parts of the parameter space. The key result and value of NRE lies in the utility of our constrained prior from round $R$. The final constrained prior, $\tilde{p}^{(R)}(\btheta)$, along with previously generated and cached samples, allows for estimation of \emph{any} higher dimensional marginal posterior $\tp^{(R)}(\bvartheta|\bx_0) \approx p(\bvartheta|\bx_0)$ of interest by doing likelihood-to-evidence ratio estimation. 
The NRE algorithm is detailed in the supplementary material.

\paragraph{Inhomogeneous Poisson Point Process (iP3) Sample Caching.}
Simulating $(\bx, \btheta)\sim p(\bx|\btheta)p(\btheta)$ can be extremely expensive. We develop a scheme to systematically reuse appropriate subsets of previous simulator runs. Our method samples $N\sim \pois(\hat N)$ parameter vectors from an arbitrary distribution $p(\btheta)$, where $\hat N$ is the expected number of samples. 
Taking $N$ samples from $p(\btheta)$ is equivalent to drawing a single sample
$\Theta \equiv \{\btheta^{(n)}\}_{n=1}^{N}$
from an inhomogenous Poisson point process (PPP) with intensity function $\lambda(\btheta) = \hat{N} p(\btheta)$. In this context, $\Theta$ is known as a set of \emph{points}. This formulation provides convenient mathematical properties \cite{ppp}, at the low price of introducing variance in the number of samples drawn. 
The precise number of samples does not matter as long as $N \approx \hat{N}$, which is true in our regime of order $\geq 1000$.

We will need two properties of PPPs.  \emph{Superposition:} Given two independent PPPs with intensity functions $\lambda_{1}(\btheta)$ and  $\lambda_{2}(\btheta)$, the sum yields another PPP with intensity function $\lambda(\btheta) = \lambda_{1}(\btheta) + \lambda_{2}(\btheta)$. 
The union of two sets of points  $\Theta = \Theta_1 \cup \Theta_2$ from the individual PPPs is equivalent to a single set of points from the combined PPP.
\emph{Thinning:} Consider a PPP with intensity function $\lambda(\btheta)$, and an arbitrary function $q(\btheta): \mathbb{R}^{d} \to [0, 1]$. 
If we are interested in drawing from a PPP with intensity function $\lambda_{q}(\btheta) = q(\btheta) \lambda(\btheta)$, we can achieve this by drawing a set of points $\Theta$ distributed like $\lambda(\btheta)$ and then rejecting individual points $\btheta^{(n)}$ with probability $1 - q(\btheta^{(n)})$.

We define a parameter cache by a set of points $\Theta_{sc}$ drawn from a PPP with intensity function $\lambda_{sc}(\btheta)$.  For every point $\btheta\in\Theta_{sc}$, a corresponding observation $\bx$ is stored in an observation cache $\mathcal{X}_{sc}$.
Our iP3 cache sampling algorithm that is responsible for maintaining the caches and sampling from a PPP with target intensity function $\lambda_t(\btheta) = \hat{N} p(\btheta)$ is written out in the supplementary material.
It is summarized in two steps: First, consider all points $\btheta \in \Theta_{sc}$ from the cache and accept them with probability 
$\min(1, \lambda_t(\btheta)/\lambda_{sc}(\btheta))$.
The thinning operation yields a sample $\Theta_1$ from a PPP with intensity function 
$\lambda_1(\btheta) = \min(\lambda_t(\btheta), \lambda_{sc}(\btheta))$. Second, draw a new set of points $\Theta_p$ from $\lambda_t(\btheta)$, and accept each $\btheta\in\Theta_p$ 
with probability $\max(0, 1-\lambda_{sc}(\btheta)/\lambda_t(\btheta))$.  This yields a sample $\Theta_2$ from a PPP with intensity function $\lambda_2(\btheta) = \max(0, \lambda_t(\btheta) - \lambda_{sc}(\btheta))$.  Thanks to superposition, the union $\Theta_1 \cup \Theta_2 = \Theta_t$ yields a sample from the PPP with intensity function $\lambda_t(\btheta)$--the sample we were looking for. We only need to run simulations on points from $\Theta_1$. Points in $\Theta_2$ already have corresponding observations in $\mathcal{X}_{sc}$ which we can reuse. Finally, the new parameters are appended to the set of points in the parameter cache, $\Theta_{sc} \to \Theta_{sc} \cup \Theta_2$. Similar for $\mathcal{X}_{sc}$. On the basis of the superposition principle, the intensity function of the $\Theta_{sc}$ cache is updated
$\lambda_{sc}(\btheta) \to \max(\lambda_{sc}(\btheta), \lambda_t(\btheta))$.

Storing and updating the parameter cache's intensity function $\lambda_{sc}(\btheta)$ can pose challenges when it is complex and high-dimensional. Our NRE implementation overcomes these challenges by learning marginal 1-dim posteriors, guaranteeing that the relevant target intensities always factorize, $\lambda_t(\btheta) = \lambda_t(\theta_1)\cdots \lambda_t(\theta_d)$. Storage of and calculation with factorizable functions simplifies matters.

\section{Experiments and discussion}

Although NRE is applicable to scenarios where the likelihood is intractable, we focus on examples with tractable likelihoods in order to compare our results with the ground truth obtained by sampling techniques. In this case, ground truth results are provided by MultiNest~\cite{Feroz2008, pymultinest}, 
a nested sampling tool widely used in the physics and astronomy community.  Examples are of illustrative nature, a quantitative comparison between methods will be left for future study.

\begin{figure}[h]
    \centering
   \includegraphics[width=0.47\textwidth]{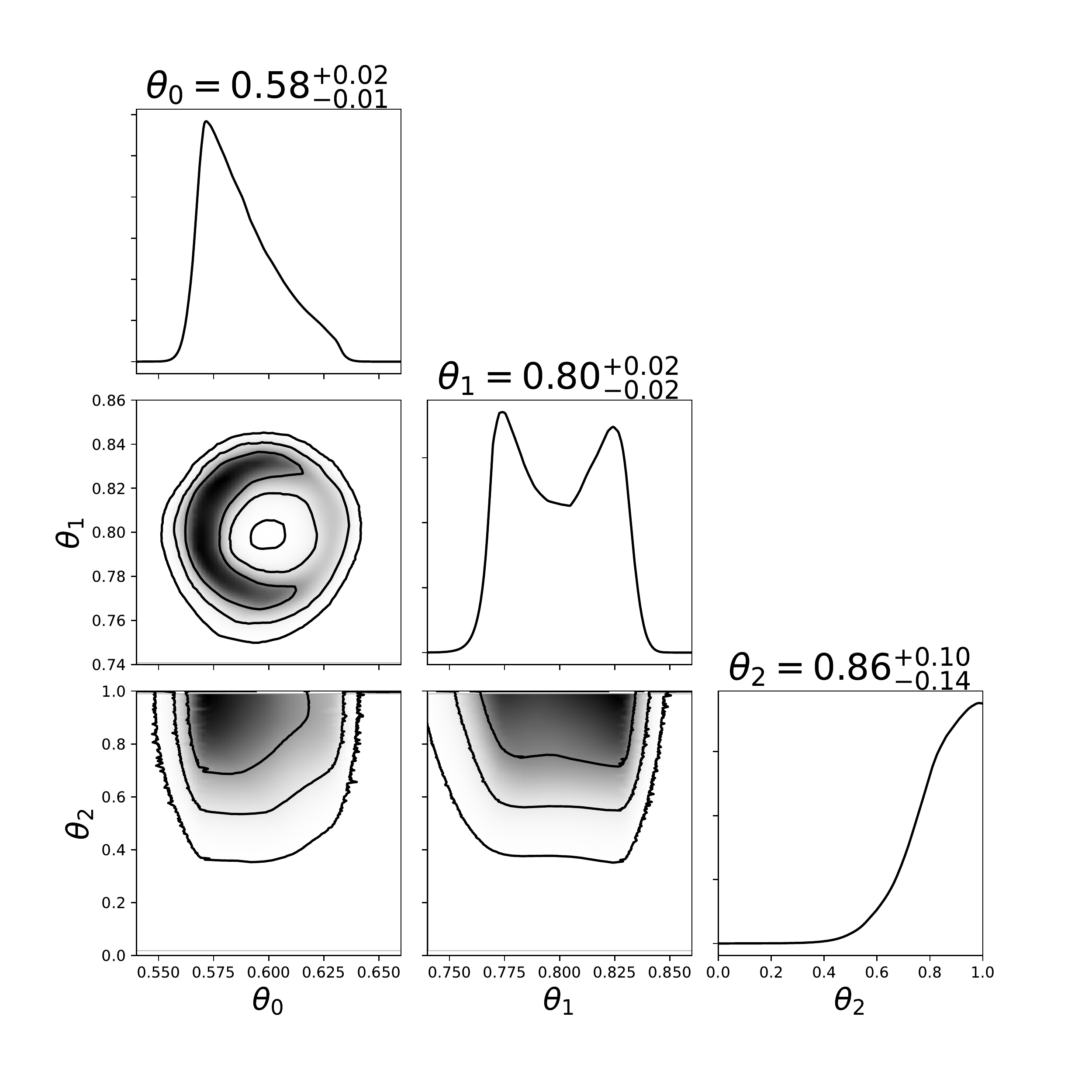}\hfill \includegraphics[width=0.45\textwidth]{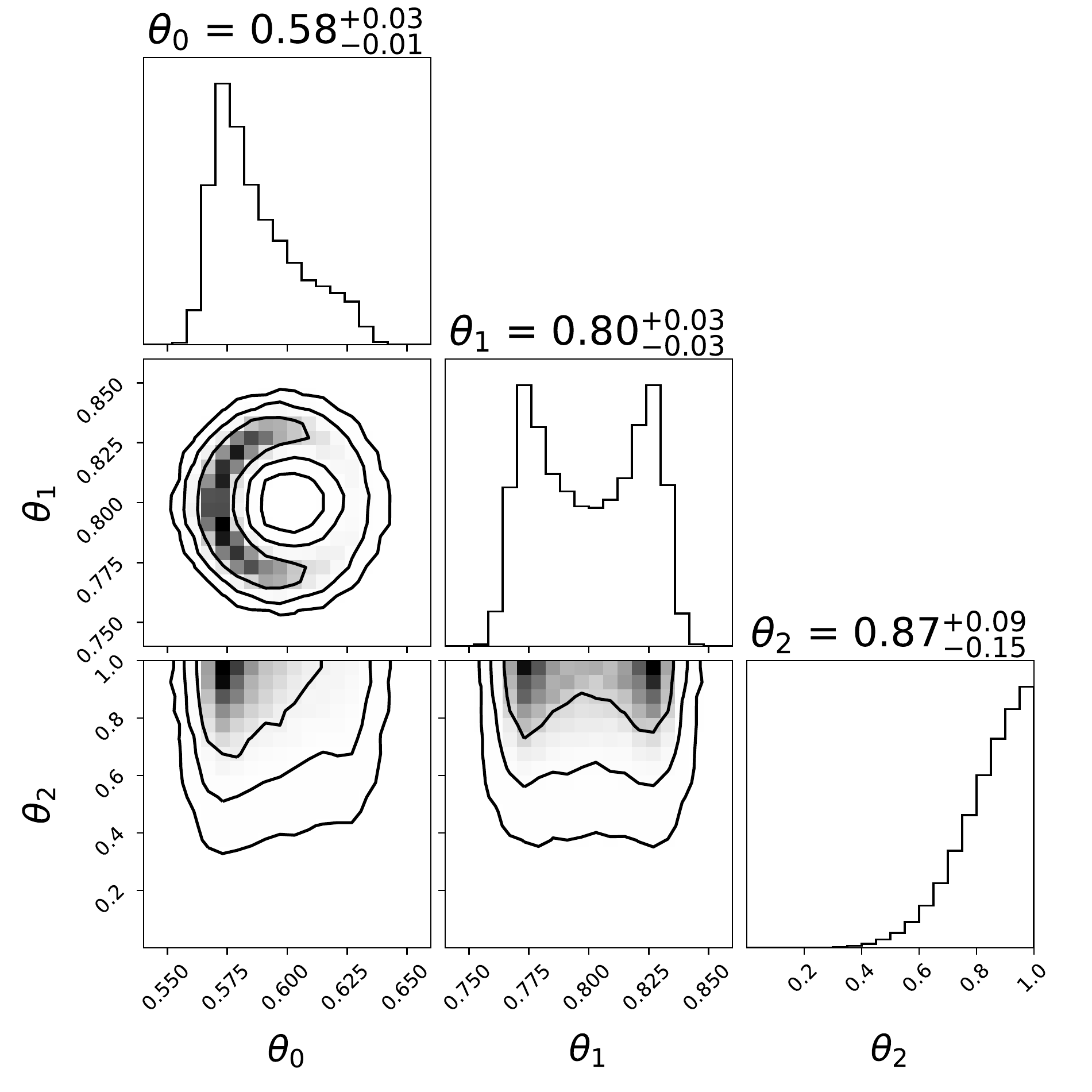}
    \caption{Marginal posteriors produced by NRE/iP3 (left) and MultiNest (right) for Run 1 of our example. The contours for $68\%,~95\%,$ and $99.7\%$ credible regions are shown.}
    \label{fig:trianglePlots}
\end{figure}

\paragraph{Experiments.} Consider a simulator $g(\btheta, \bz) = (\theta_0,\sqrt{(\theta_0-0.6)^2+(\theta_1-0.8)^2},\theta_2)+\textbf{n}$, where $\textbf{n}$ is drawn from a zero-mean multivariate Gaussian distribution with a diagonal covariance matrix,  $\boldsymbol{\Sigma}=\textrm{diag}(0.03, 0.005, 0.2)$. The first task was to infer $\btheta$ for synthetic data generated by ground truth parameters $\btheta=(0.57, 0.8, 1.0)$. In our reported estimates, as seen in Fig.~\ref{fig:trianglePlots}, MultiNest required 160722 simulations, while NRE obtained comparable marginal posteriors after only 20011 simulations divided across four rounds. A visualization of the cache $\Theta_{sc}$, utilized in NRE over four rounds, is presented in the supplementary material (see Fig.~ \ref{fig:cacheContents}). The second task was to infer new parameters $\btheta=(0.55, 0.8, 1.0)$ with the same simulator. MultiNest required 171209 new simulator runs while iP3 sample caching reduced the number of additionally required simulator runs to only {3668} when performing NRE. NRE is efficient initially and enables hyper-efficient follow up studies. 

\begin{figure}
    \centering
    \includegraphics[width=0.45\textwidth]{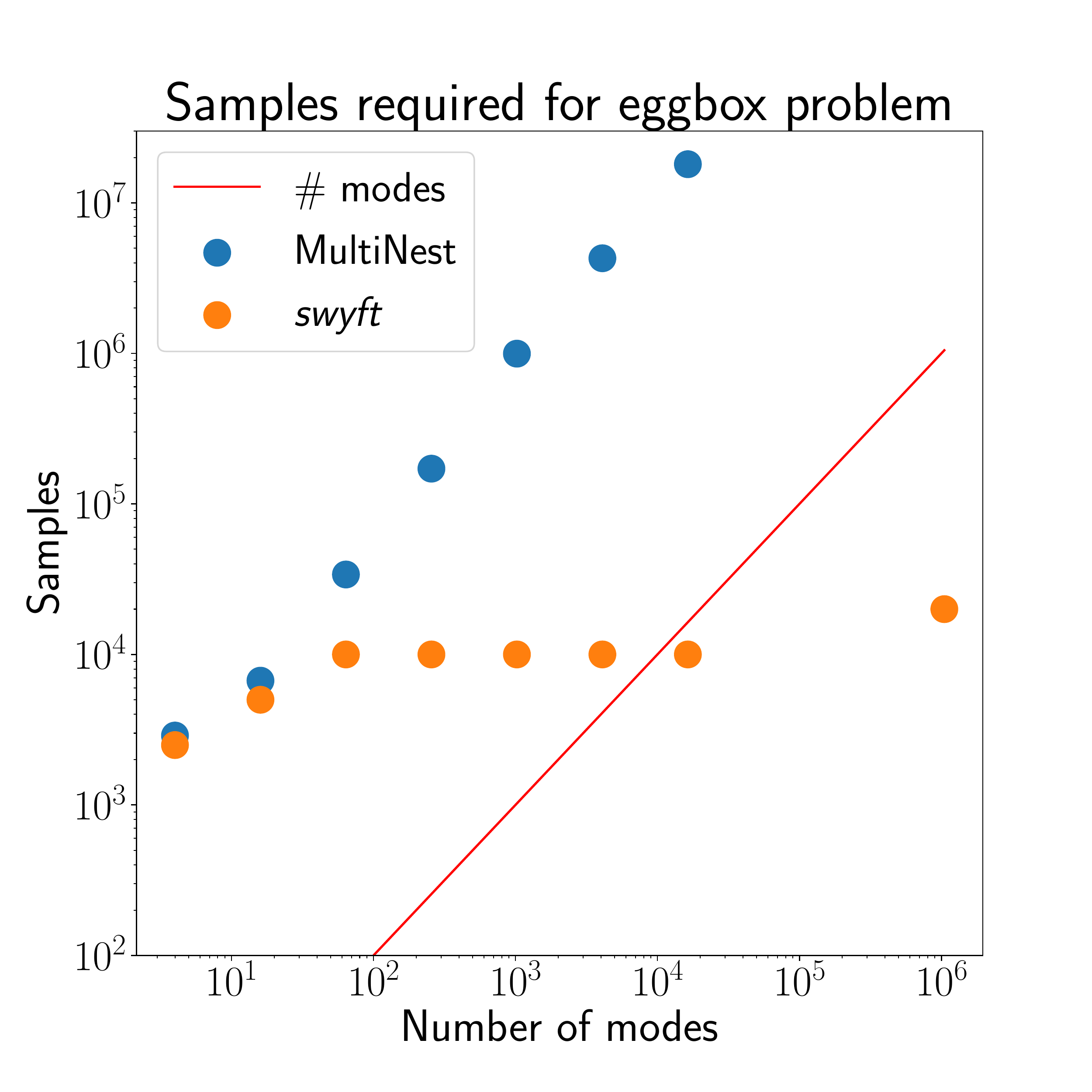}\includegraphics[width=0.45\textwidth]{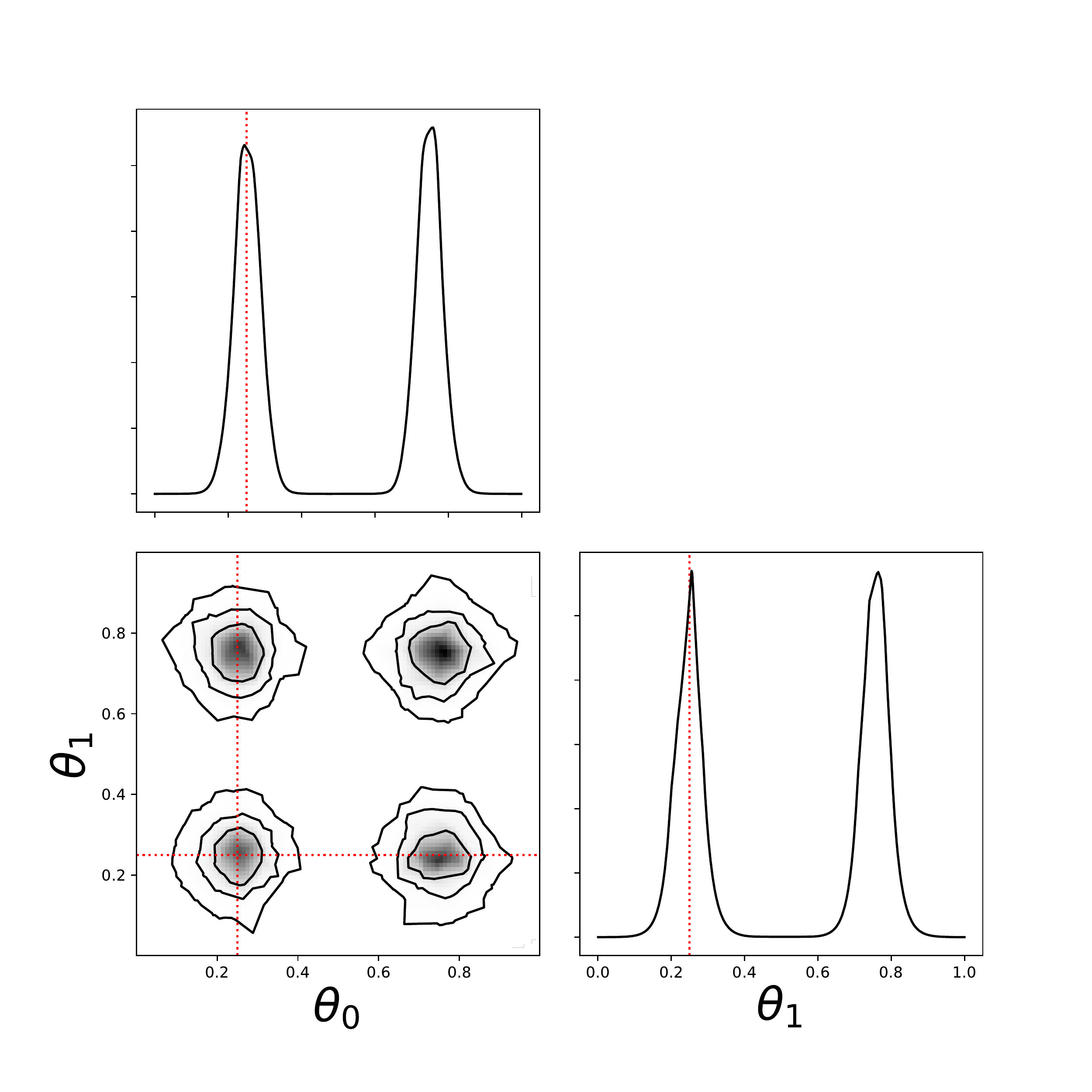}
    \caption{
    Left: scaling of simulator cost with respect to number of modes for the eggbox problem. For MultiNest, each mode must be sampled, leading to a rapid increase in simulator cost as a function of problem dimension.  On the other hand, by directly evaluating the marginal posteriors, the proposed method is able to solve the problem at almost constant computational cost.  Right: posteriors computed via a network for $d=20$ trained on $2\times 10^4$ samples. For ease of visualization, we show only the set of 1- and 2-dimensional marginal posteriors truncated to $\theta_0,\theta_1$; marginal posteriors for all other parameter pairs look similar.
    }
    \label{fig:eggbox}
\end{figure}

In our second experiment, the simulator gives rise to an ``eggbox'' posterior, with modes at $\theta_i=0.5\pm 0.25$, $i=0,\dots d-1$. The simulation model is $g_i(\boldsymbol{\theta}, \mathbf{z}) = \sin(\theta_i \cdot \pi) + n_i$, where $\mathbf{n}$ is a zero-mean Gaussian noise with standard deviation 0.1.  For this model, the number of modes grows exponentially with the dimension of $\btheta$ as $2^{d}$. As such, the number of samples required for MultiNest scales exponentially with the dimension and quickly becomes infeasible, see Fig.\ \ref{fig:eggbox}. At $d=20$ there are over $10^6$ modes and MultiNest cannot solve the problem before memory constraints take over. We find that, with standard settings, MultiNest requires at least $10^{7}$ samples for $d=14$ to give reasonable results (which corresponds to $\sim 10^3$ samples per mode).  On the other hand, our proposed method \emph{directly estimates marginal} posteriors at almost constant simulator cost, with $2\times 10^4$ samples sufficient even for $d=20$ (much less than one sample per mode). Since we focus with NRE directly on marginal posteriors, resolving the exponentially large number of modes becomes unnecessary.

\begin{table}[h]
    \floatbox[\capbeside]{table}[0.4\textwidth]{%
        \caption{Numbers of samples required by NRE/iP3 and MultiNest for first experiment. We infer posteriors for two instances of the problem. In the second instance, NRE/iP3 benefits from sample reuse while MultiNest cannot reuse any parameters or simulations.}
        \label{tab:test}
    }{%
    \begin{tabular}{ccc}
        \toprule
         & \bf{NRE/iP3} & \bf{MultiNest}\\
        \midrule
        Run 1 & 20011 & 160722  \\
        Run 2 & {3668} & 171209\\
        \bottomrule
    \end{tabular}\\
    }
\end{table}

\paragraph{Discussion.}
NRE with iP3 sample caching is meant as a significantly more flexible, and simulator efficient, alternative to the likelihood-based sampling tools commonly used in physics.   In the above experiment we show that -- without much tuning -- our algorithms achieved similar accuracy to the widely used nested sampler MultiNest while reducing simulator calls by an order of magnitude. Furthermore, analysis of similar inference problems with usual sampling tools, like MultiNest, must start from scratch and redo many simulations; however, NRE with iP3 sample caching allows us to reuse simulator runs from previous analyses to further reduce simulation costs. NRE is applicable to simulators which do not have a tractable likelihood, significantly broadening the set of use-cases. 

In our reference implementation of NRE with iP3 sample caching, the initial rounds are based on simultaneous learning of one dimensional marginal posteriors, implemented efficiently via multi-target training.  Once converged, higher-dimensional marginals can be scrutinized in the relevant region after training a new likelihood ratio estimator on already-collected data. Empirically, using low dimensional marginals and constraining the prior to relevant regions of parameter space reduces the difficulty of the learning problem such that relatively simple neural networks are sufficient to estimate satisfactory marginal posteriors. In future work we will consider how to further capitalize on this learning problem simplification by using higher dimensional marginal priors and posteriors during the training rounds of NRE. The aim is to improve prediction across highly correlated parameters. 
\emph{As a product of these efforts, we offer an open source Python package which implements NRE and iP3 sample caching called \textbf{swyft}. It has a convenient API for use by machine learning non-specialists.}

\section*{Broader Impact}
The proposed algorithms aim at facilitating solutions to the "inverse problem" for simulator-based modeling in the limit of expensive simulator runs. Although we have physics and astronomy applications in mind, the underlying problem is rather common in the quantitative sciences. For example, benchmark problems include the Lotka–Volterra predator-prey model and the m/d/1 queue problems. Therefore, the breadth use cases must be considered and the software may be used to make modeling decisions which affect the lives of humans.
We do not anticipate specific cases of misuse of the methods, but emphasize that the inference method is only as good as the simulator, and limited simulation models can lead to false conclusions.
A problematic failure mode would be to choose a simulator which reinforces already held biases. Our method does not absolve the scientist of the responsibility on the choice of simulator.
However, our algorithms do not directly penalize model complexity, and hence allow and encourage the construction of more realistic models.

\begin{ack}
This work uses \texttt{numpy} \cite{harris2020array}, \texttt{scipy} \cite{2020SciPy-NMeth}, \texttt{matplotlib} \cite{Hunter:2007}, \texttt{pytorch} \cite{pytorch}, and \texttt{jupyter} \cite{jupyter}. Benjamin Kurt Miller is funded by the University of Amsterdam Faculty of Science (FNWI), Informatics Institute (IvI), and the Institute of Physics (IoP).
This work is part of a project that has received funding from the European Research Council (ERC) under the European Union’s Horizon 2020 research and innovation programme (Grant agreement No. 864035 -- UnDark).

\end{ack}

\bibliographystyle{unsrt}
\bibliography{bibliography}

\clearpage

\section*{Supplementary Material}

\begin{algorithm}[H]
\DontPrintSemicolon
\SetAlgoLined
\SetKwBlock{KwProb}{with}{end}
\SetKwInOut{Input}{Input}
\SetKwInput{Init}{Init}
\SetKwFunction{ippp}{iP3}
\Input{Simulator $p(\bx \vert \btheta)$, factorizable prior $p(\btheta)$, real observation $\bx_{0}$, parameter dimension $d$, classifiers $\{f_{\phi, i}(\bx, \theta_{i})\}_{i=1}^{d}$, rounds $R$, mean samples per round $\hat{N}$, likelihood cutoff $\epsilon$.}
\Init{Constrained prior $p^{(1)}(\btheta) = p(\btheta)$, caches $\mathcal{X}_{sc}, \Theta_{sc}= \{ \}$, intensity function $\lambda_{sc}(\btheta) = 0$.}
\For{$r=1$ \KwTo $R$}{
    Sample points using Algorithm \ref{alg:ip3}, the iP3 sample caching algorithm, \\
    $\{(\bx^{(n)}, \btheta^{(n)})\}_{n=1}^{N \sim \pois(\hat{N})}$,
    $\mathcal{X}_{sc}$, 
    $\Theta_{sc}$, 
    $\lambda_{sc}(\btheta)$ 
    = \ippp$(p(\bx \vert \btheta), \hat{N}, p^{(r)}(\btheta), \mathcal{X}_{sc}, \Theta_{sc}, \lambda_{sc}(\btheta))$.\;
    (Re-) initialize $\mathbf{f}_{\phi} = (f_{\phi, i})_{i=1}^{d}$. \\
    \While{$\mathbf{f}_{\phi}$ not converged}{
    	Extract mini-batch $\{(\bx^{(b)}, \btheta^{(b)})\}_{b=1}^{B} \subset \{(\bx^{(n)}, \btheta^{(n)})\}_{n=1}^{N \sim \pois(\hat{N})}$, $B \equiv 0 \pmod{2}$.\;
	    Randomly pair up all samples from the mini-batch $((\bx^{(a, 0)}, \btheta^{(a, 0)}), (\bx^{(a, 1)}, \btheta^{(a, 1)}))$ where $a=1, \dots, B/2$.\;
	    Minimize 
	    $\mathcal{L}(\phi) 
	    = -\frac{1}{B} \sum_{a=1}^{B/2} \sum_{i=1}^{d} \sum_{j \in \{0, 1\}}
	    \sigma(f_{\phi, i}(\bx^{(a, j)}, \theta_{i}^{(a, j)})) +
	    \sigma(-f_{\phi, i}(\bx^{(a, j)}, \theta_{i}^{(a, \neg j)}))
	    $ using stochastic gradient descent or a variant.\;
	}
	Constrain
	$p^{(r+1)}_{i}(\theta_{i}) \propto p_{i}(\theta_{i}) \, \Theta_{H} 
	\left[ \frac{\exp(f_{\phi, i}(\bx_{0}, \theta_{i}))}{\max_{\theta_{i}} \exp(f_{\phi, i}(\bx_{0}, \theta_{i}))} - \epsilon \right]$ in order to construct constrained prior 
	$p^{(r+1)}(\btheta) 
	= p^{(r+1)}_{1}(\theta_{1}) \cdots p^{(r+1)}_{d}(\theta_{d})$.\;
}
\caption{Nested Ratio Estimation (NRE)}
\label{alg:nre}
\end{algorithm}

\begin{algorithm}[H]
	\SetKwInOut{Input}{Input}
	\SetKwInOut{Output}{Output}
	\SetKwInput{Init}{Init}
	\DontPrintSemicolon
	\SetAlgoLined
	\SetKwBlock{KwProb}{with}{end}
	\Input{Simulator $p(\bx \vert \btheta)$, mean samples per round $\hat{N}$, constrained prior $p(\btheta)$, \\
	observation cache $\mathcal{X}_{sc}$, parameter cache $\Theta_{sc}$, intensity function $\lambda_{sc}(\btheta)$.\;}
	\Output{Samples $\{(\bx^{(n)}, \btheta^{(n)}) : (\bx^{(n)}, \btheta^{(n)}) \sim p(\bx \vert \btheta) p(\btheta) \, \forall n\}_{n=1}^{N \sim \pois(\hat{N})}$, \\
	observation cache $\mathcal{X}_{sc}$, parameter cache $\Theta_{sc}$, parameter cache intensity function $\lambda_{sc}(\btheta)$.\;}
	\Init{Number of points $N \sim \pois(\hat{N})$, size of cache $M = \vert \Theta_{sc} \vert\ = \vert \mathcal{X}_{sc} \vert$, output set $\mathcal{O} = \{\}$, target intensity function $\lambda_{t}(\btheta) = \hat{N} p(\btheta)$.\;}
	\For{$m = 1$ \KwTo $M$}{
		\KwProb(probability {$\min(1, \lambda_t(\btheta)/\lambda_{sc}(\btheta))$} \textbf{do}){
			Get observation by index $\bx^{(m)} = \mathcal{X}_{sc}^{(m)}$.\;
			Get parameter vector by index $\btheta^{(m)} = \Theta_{sc}^{(m)}$.\;
			Append sample to output set $\mathcal{O} = \mathcal{O} \cup \{(\bx^{(m)}, \btheta^{(m)})\}$.\;
		}
	}
	\For{$n = 1$ \KwTo $N$}{
		Draw parameter sample $\btheta^{(n)} \sim p(\btheta)$.\;
		\KwProb(probability {$\max(0, 1-\lambda_{sc}(\btheta)/\lambda_t(\btheta))$} \textbf{do}){
			Simulate $\bx^{(n)} \sim p(\bx \vert \btheta^{(n)})$.\;
			Append sample to output set $\mathcal{O} = \mathcal{O} \cup \{(\bx^{(n)}, \btheta^{(n)})\}$.\;
		}
	}
	Update parameter cache $\Theta_{sc} = \Theta_{sc} \cup \{\btheta^{(n)}\}_{n=1}^{\vert \mathcal{O} \vert \sim \pois(\hat{N})}$.\;
	Update observation cache $\mathcal{X}_{sc} = \mathcal{X}_{sc} \cup \{\bx^{(n)}\}_{n=1}^{\vert \mathcal{O} \vert \sim \pois(\hat{N})}$. (Note, both caches are sets indexed by the order in which elements were added to them. Duplicate elements are ignored.)\;
	Update intensity function of parameter cache $\lambda_{sc}(\btheta) = \max(\lambda_{sc}(\btheta), \lambda_{t}(\btheta))$.\;
	\Return $\mathcal{O}$, $\mathcal{X}_{sc}$, $\Theta_{sc}$, $\lambda_{sc}(\btheta)$.\;
	\caption{Inhomogeneous Poisson Point Process (iP3) Sample Caching}
	\label{alg:ip3}
\end{algorithm}

\begin{figure}
    \centering
    \includegraphics[width=\textwidth]{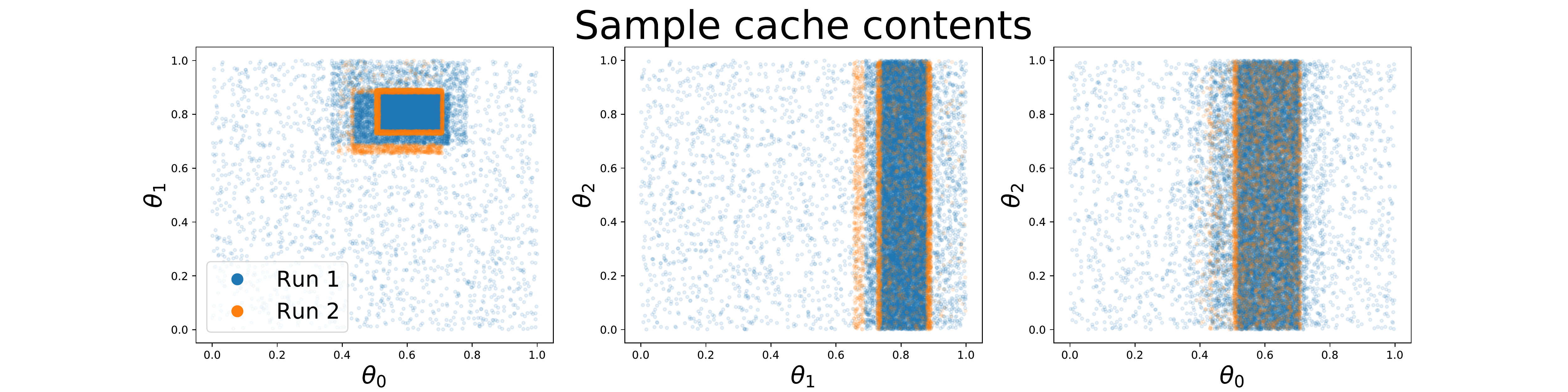}
    \caption{Parameter cache $\Theta_{sc}$ contents for our first experiments. (See Table \ref{tab:test}.) Samples added in the first and second runs are shown in blue and orange, respectively. The second run reuses samples computed during the first run. The evolution of the cache towards relevant parameter regimes is visible. }
    \label{fig:cacheContents}
\end{figure}

\end{document}